# Investigating Asteroid Surface Geophysics with an Ultra-Low-Gravity Centrifuge in Low-Earth Orbit

Stephen R. Schwartz[a,*], Jekan Thangavelautham[b], Erik Asphaug[a], Aman Chandra[b], Ravi teja Nallapu[b], Leonard Vance[b]

[a] *Lunar and Planetary Laboratory, University of Arizona, Tucson, Arizona, United States 85721*, srs@lpl.arizona.edu
[b] *Space and Terrestrial Robotics Laboratory (SpaceTREx), Department of Mechanical and Aerospace Engineering, University of Arizona, Tucson, Arizona, United States, 85721.*
\* Corresponding Author

## Abstract

Near-Earth small-body mission targets 162173 Ryugu, 101955 Bennu, and 25143 Itokawa produce gravity fields around 4 orders of magnitude below that of Earth and their irregular shapes, combined with rotational effects produce varying surface potentials. Still, we observe familiar geologic textures and landforms that are the result of the granular physical processes that take place on their surfaces. The nature of these landforms, however, their origins, and how these surfaces react to interrogation by probes, landers, rovers, and penetrators, remain largely unknown, and therefore landing on an asteroid and manipulating its surface material remains a daunting challenge. The AOSAT+ design is a 12U CubeSat that will be in Low-Earth Orbit (LEO) and that will operate as a spinning on-orbit centrifuge. Part of this 12U CubeSat will contain a laboratory that will recreate asteroid surface conditions using crushed meteorite as a regolith proxy. The spinning of the laboratory will simulate the surface gravity of asteroids 2 km and smaller. The result is a bed of realistic regolith, the environment that landers and diggers and maybe astronauts will interact with. A crucial component of this mission involves the reproduction of the experimental results in numerical simulation in order to extract the material parameters of the regolith and its behavior in a sustained, very low—but nonzero—gravity environment.

## 1. Introduction

There are tens of thousands of asteroids larger than 100 m in near-Earth space and millions more in the Main Belt [1]. Their physical properties and compositions are diverse, and they hold secrets to the early Solar System epoch as minimally processed remnants of planetary formation. This makes them valuable for planetary science and strategic for resource mining. An understanding of their properties is also essential for detailed exploration and to the development of mitigation strategies against asteroids that pose a risk of impacting the Earth. However, what we lack is direct knowledge of the geophysical behavior of the asteroid surface, especially under milligravity conditions.

In recent decades, radar-imaging capabilities have improved and spacecraft have been sent to image these small bodies on flybys and rendezvous missions. We have learned a great deal about their surface geologies in terms of regolith cover, boulder and crater populations, and near-surface structure (scarps, saxons, and outcroppings). We have now entered an age where we have begun to interface with the surfaces of these bodies, retrieving samples, depositing landers, and performing impact experiments (Deep Impact, Hayabusa2, DART [2-4]).

Marking the end of NASA's successful discovery-class mission, the Near Earth Asteroid Rendezvous – Shoemaker (NEAR Shoemaker) was the first spacecraft to soft-land (touchdown) on an asteroid (433 Eros) [5]. In 2010, JAXA's Hayabusa spacecraft, after surveying in great detail the small, peanut-shaped asteroid 25143 Itokawa, returned a tiny sample from its surface back to Earth. Beginning in late 2018, JAXA's Hayabusa2 mission has surveyed asteroid 162173 Ryugu, deployed multiple landers onto the asteroid, collected two samples from its surface, and performed a 2 km/s-impact experiment to produce an artificial crater and to reveal subsurface material. NASA's OSIRIS-REx mission arrived at asteroid 101955 Bennu in early December 2018 and has relayed exquisite visual images, detailed terrain maps, thermal maps, gravity data and more back to Earth and is preparing to acquire a sample from its target in July 2020 [6].

The next steps in space exploration will involve progressively more complex interactions with the surfaces of small planetary bodies. We can see this from the evolution of small body missions sponsored by NASA, JAXA, The China National Space Administration, among others. Each mission can cost hundreds-of-millions to billions of dollars, plus many years to develop and execute. The complexity and the safety of what they can accomplish is limited by our existing knowledge of how these airless, low-gravity rubble-pile surfaces reacts to robotic mechanisms. This includes the compliance of the surface to penetration by mechanical arms, landers, and rovers, and this compliance depends on cohesional and frictional properties that are largely unconstrained. Conducting inventive, inexpensive experiments targeted at uncovering these properties could teach us as much, if not more, than a billion-dollar mission and is the most prudent pathway to advancing our capabilities in interplanetary space exploration. This is both the quickest and the most cost-effective approach.

Towards this goal, we are putting forth plans for AOSAT+ [7], a 12U CubeSat that will be in Low-Earth Orbit (LEO) and that will operate as a spinning on-orbit centrifuge. This is a follow-on to AOSAT 1 [26-28] Part of this 12U CubeSat will contain a laboratory that will recreate asteroid surface conditions using crushed meteorite as a regolith proxy. The spinning of the laboratory will simulate the surface gravity of asteroids 2 km and smaller. The result is a bed of realistic regolith, the environment that landers and diggers and maybe astronauts will interact with.

JAXA's Hayabusa2 mission [3] and NASA's OSIRIS-REx mission [6] are performing the most robust to-date experiments involving regolith surfaces in sub-milligravity environments, using sampler heads, and, in the case of Hayabusa2, a small hypersonic impactor. These missions are or will be obtaining images, thermal data, and other measurements indicating the force response to surface interrogations. But OSIRIS-REx, for example, will only be interacting directly with the surface during its "touch-and-go" sample-acquisition, which may put the spacecraft in contact with the surface for about ten seconds. With AOSAT+, watching simple, but informative granular dynamics play out in a low-gravity environment for hours to days, without the ops- and risk-related constraints on data-acquisition time, will compliment and help to explain the sample-acquisition data from these small-body sampling missions. As well, it will be an early step in the development of more complex surface interactions in the future. We will exploit the benefits that come with having a long-term, sustained experiment, including the ability to observe the dynamics inside the chamber continuously for long periods—important given the low speeds and weak forces.

Our team has extensive experience in the development and use of granular codes, with validation from terrestrial experiments [8-10], but applied specifically to planetary science studies of small-body formation and regolith dynamics. One key strength that we have is the ability to numerically recreate the conditions of the AOSAT+ lab chamber with exquisite precision (using a suite of numerical tools described below), which will allow us to extract essential material parameters of the grains in the sub-milligravity environments of interest. The series of experiments to be performed aboard the AOSAT+ laboratory, described in detail below (Sec. 7), will each contribute to uncovering the material parameters of regolith dynamics in the low-energy, low-gravity regimes relevant to the interfacing between mechanisms and an asteroid or cometary surface.

## 2. Spacecraft Overview

AOSAT+ will operate as an on-orbit science laboratory to simulate the surface conditions of small-bodies such as Didymos-B, Bennu, Ryugu and Phobos. It is designed as a 12U, 24-kg CubeSat that will centrifuge by spinning about its short axis (from 0.1 up to a few rotations-per-minute). The centrifugal force will simulate the low gravity field of small asteroids.

The design of AOSAT+ includes a suite of onboard instruments to characterize the dynamics and behavior of crushed meteor particles under asteroid surface gravity conditions relevant to planetary science, asteroid robotics, and in-situ resource utilization. The details of the spacecraft components are described in [7], but include a 12U bus developed by the Space Dynamics Laboratory (SDL), two large body-mounted deployable eHawk solar panels (TRL 9) containing triple junction cells (to avoid the complexity involved in gimbaling), and a bank of 144 Whr Yardney Lithium Ion batteries (also TRL 9). The spacecraft will also use an SDL-built Single Board Computer containing a Cobham Leon 3FT rad-hardened processor (TRL 9) operating at 125 Mhz coupled with a rad-tolerant Micro-semi FPGA, 256 MB of SDRAM, and a 16-GB solid state drive. Via SpaceWire, the SBC board interfaces to the Mission Unique Card (MUC), which hosts all the science instruments. The SBC is rad-tolerant, thus enabling robust handling of Single Event Upsets (SEUs) and Random Bit Flips (RBFs). The proposed computer board has been successfully combined with the proposed Innoflight SCR-101 S-band Radio (TRL 8). The design also includes four Raspberry PI 3 Compute Modules (TRL 9) that act as a controller for each instrument, and are connected to the MUC card. The on-board Attitude Determination Control System (ADCS) is handled by the Blue Canyon Technologies XACT-50.

## 3. Instruments

AOSAT+ is configured with cameras, lasers, actuators, and small mechanical instruments to both observe and manipulate the regolith at low simulated gravity conditions [7]. A series of experiments at varying effective gravity will measure the general behavior, internal friction, adhesion, dilatancy, coefficients of restitution, and other parameters that will be modeled and constrained by extensive numerical simulations (Sec. 5).

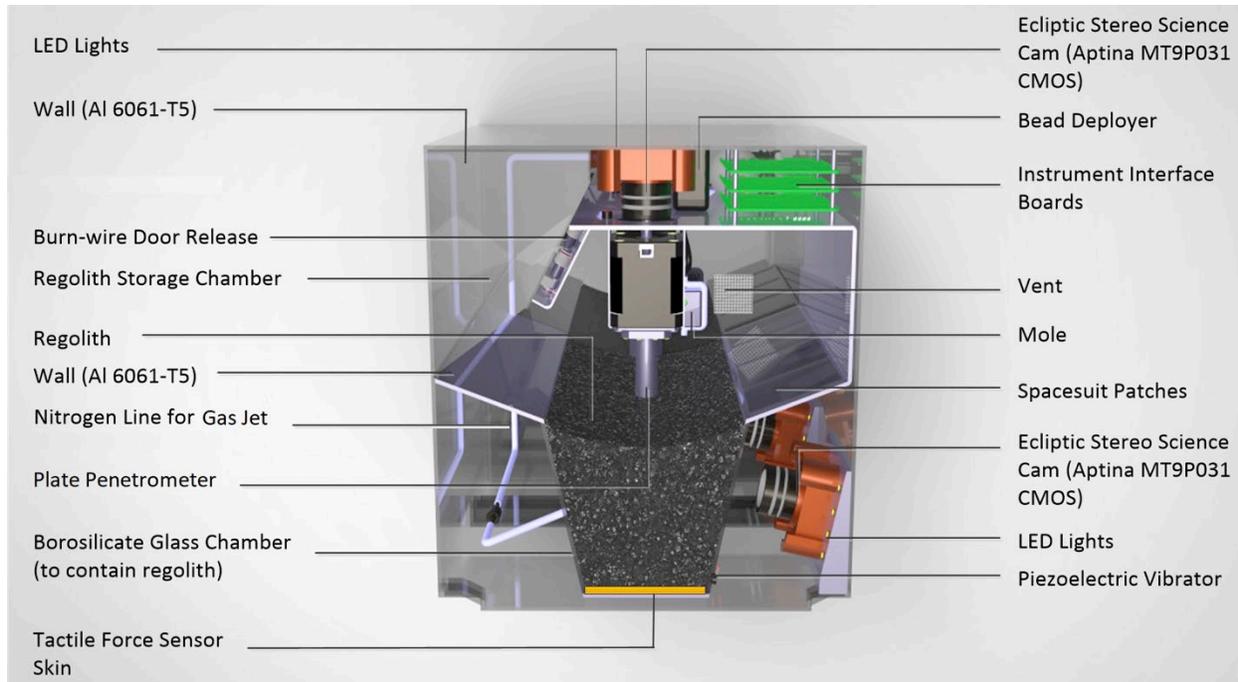

**Fig. 1**. The AOSAT+ lab chamber is housed in the lower two-thirds of the spacecraft away from the flight avionics. This configuration maximizes the centripetal radius to attain maximal-g for lowest spacecraft spin (angular velocity). The chamber is shown in Figure 7. Regolith (black) is shown in its stowed configuration; the door drops and the regolith comes out, aided by centrifugal force and vibrations. Shown are the spacesuit material coupons, the tactile skin, optical systems (includes cameras, LEDs and laser sheets), and actuators (including vibrators, Gas Jet, bead deployer). The mole (SEO) will be mounted inside the regolith stowage chamber, for deployment after Baseline EOM.

The AOSAT+ lab chamber, shown in Fig. 1, occupies more than half of the 12U CubeSat, isolated from the flight avionics. Shown are the spacesuit material coupons, the tactile skin, optical systems (includes cameras, LEDs and laser sheets), and actuators (including vibrators, Gas Jet, bead-deployer).

### 4. Experiments

AOSAT+ is a fundamental science investigation whose broad objective is to make substantial, quantitative inroads into the underlying regolith physics of asteroids, comets and small moons. This requires acquiring direct knowledge, in the milligravity environment of asteroids, about the _surface hardness_ and regolith response/compliance to interrogation by a robotic platform; the _granular dynamics_ in ultra-low (but non-zero) directional gravity fields; _gas entrainment_ as volatiles interact with and permeate through loose regolith materials; _ballistic penetration_ of particles striking the surface at speeds comparable to escape speed; _material interactions_ between native regolith and spacesuit fabrics, metallic surfaces, and glass surfaces, and the effects of vibrations and gas jets on dust mitigation.

To study each of these aspects, AOSAT+ is equipped with a _plate penetrometer_, _vibrators_, a _gas jet_, _bead deployer_ and four coupons of _spacesuit fabric_, respectively.

A force-bearing plate penetrometer will measure the strength of the regolith surface with a linear actuator rod and bi-directional force sensor. The resisting force will be recorded (at 100 Hz) as the actuator extends slowly (quasistatically) from above the regolith to within ~1 cm (size of the largest particles) from the bottom. The design is inspired by the slow mode of penetration of comet penetrometers [11]. Besides providing a direct measurement of the compliance of the regolith, crucial in landing a spacecraft or in penetration operations, since this experiment is designed to maintain a quasi-static state, it largely isolates out many of the plastic damping parameters. This will thus provide data on the elastic parameters, such as static friction and the elastic component of rolling friction (Sec. 5.1). The adhesive force will also influence these measurements.

Vibrations provide simple, solid state regolith actuation. Regolith inside a 12U CubeSat rotating at 1

RPM will be easily lofted by the simple, low-power actuation of a vibrator. There will be 12 vibrators in all, tunable to much lower energies and capable of small haptic pulses. Mimicking the dynamical response to activation in simulation will provide an incredible amount of information on the material properties and behavior at low-g. Although the effective gravity is stratfied (in the centrifuging lab space, the effitive gravity felt by a grain increases the closer it gets to the bottom of the chamber), this stratification will be explicitly modeled in simulation and material properties will be extracted from this data (Sec. 5.2). In fact, simply measuring the settling time of the grains in the chamber gives strong constraints on properties like the coefficient of restitution in the direction normal to grain contacts. This quantity, along with tangential restitution and cohesive properties, can also be extracted by observing individual collisions during settling.

A 17 cm$^3$ tank of N$_2$ gas at 0.006 bar starting pressure is used to emit controlled pulses through an in-line latching valve to provide repeated inert gas (N$_2$) interaction with regolith. Nitrogen is being used as the driver for OSIRIS-REx sample acquisition [6], and its legacy heritage has already been exploited in proposed sample return missions [12]. Matching numerical simulations of the regolith response to gas injection will be applied to simulations of the OSIRIS-REx TAGSAM operation in order to help recreate the conditions of the historic upcoming sampling event on Bennu (Sec. 5.3).

The bead deployer will be used to study higher energy ballistic interactions between particles, we use a 'pinball deployer' that can launch ten 2.5-mm aluminum spheres at a velocity of 10–100 cm/s to strike the regolith and form a crater. The baselined design of a pneumatic bead deployment system utilizes controlled bursts of low-pressure N$_2$ as a carrier for each bead. Control is affected through an in-line latching valve that opens and closes via an electrical control signal. Outside the quasistatic regime, this experiment will test the dissipation of energy within the granular medium (Sec. 5.4). Ultra-sensitive tactile force sensors, consisting of a 142-taxel 'skin' lining the regolith bed, will be developed [13]. Each skin will be capable of measuring the normal force and 2D shear force at 100 Hz at a 1-cm spatial resolution. This will provide measurements on seismic propogation and attenuation within a granular medium in a sub-milligravity regime. These quantities will address many unresolved questions in the field of planetary science including crater erasure, resurfacing, and grain-lofting on small bodies, which will affect age estimates of these bodies and their implications on Solar System formation. Using numerical models, we can pull out these parameters and apply them to simulations of asteroid surfaces (such as Bennu and Ryugu).

Thermometers and IMUs are included in the bus. IMU data will be vital for inertial reference in connecting 3D models with the movies of granular flow. Indeed, as alluded to earlier in this section, for the purpose of developing realistic regolith models, it is not necessary to have a 'perfect' centrifuge, only accurate knowledge of the state of rotation.

## 5. Numerical Methodology and Discussion

The majority of the numerical work will be carried out using the $N$-body numerical integrator PKDGRAV, a parallel gravity tree code [14] that computes collisions between spherical particles [15], and which has been well tested in regimes applicable to planetary science (e.g., [10,16]). The code uses a straightforward second-order leapfrog scheme for the integration and computes gravity moments from tree cells to hexadecapole order. Particles are considered to be finite-sized spheres and contacts are identified each step using a fast neighbor-search algorithm also used to build the tree. The code was adapted to treat hard-sphere collisions for planetesimal modeling [15] and later for granular material modeling [8] by implementing a soft-sphere discrete element method (SSDEM) collisional routine based on some of the methods described in Cundall and Strack [17].

The code we use historically treats particles as perfect spheres. Various parameters (c.f. [8]) describe their softness and plasticity, including energy storage and energy loss under compression. Other parameters describe these effects in relation to tangential stresses on their surfaces, with normal/tangential elasticity often related by a factor of 2/7 (c.f. [18,8]. Other surface effects including static friction, rolling friction, and twisting friction have been added to the code [8], with elastic deformations added to rolling and twisting friction in [19]. Many of the macroscopic consequences of nonspherical shapes have been successfully reproduced including avalanche runoff behavior (e.g., [20]) and other surface coupling effects in granular fluids. This treatment of shape effects is fast—it uses the simplicity involved in searching for contacts between spheres, with the added computational and bookkeeping expense of sophisticated friction effects by effectively averaging shape effects into spherically symmetric properties of grains—and is typically sufficient to describe macroscopic effects of realistic grain shapes, while allowing for higher resolution because of its low CPU-expense. Still, it does not treat the effects of physically interlocking grains, which depend on peculiar relations between grains and their orientations. To address this, we can "glue" spheres of different sizes together rigidly [21] or allow them to flex and bend [9] to achieve non-spherical shapes. We can also model arbitrarily shaped and arbitrarily sized individual polyhedra interacting with spherical particles.

*5.1 Plate penetrometer and surface hardness*

The precise shape, positioning, and rate of motion of the plate penetrometer onto and into the regolith bed will be simulated. This shall be performed for each experiment, and parameters will be tuned to match the results. These results include the force measurements on the plate as well as the pressure on the lab chamber boundaries as recorded by the tactile skin. Up to the Coulomb limit of the skin sensors, tangential forces will also be recorded. Matching tangential forces in simulation will provide constraints on the shearing strength between the walls and the grains, but also on indirect macroscopic inter-grain shearing resistance. Grain motion should be minimal as we approach the quasistatic limit, but information on the reorganization of the intergranular force chains the contact lattice is also a strong constraint these material properties. One of the numerical tools included in the PKDGRAV package performs a Voronoi Tesselation [19]. With this we can quantify the packing efficiency and how it evolves in low gravity starting with the settling process and continuing through induced confining pressure provided by the plate penetrometer. The most obvious application to come from this analysis will be to aid in lander design by analyzing surface hardness. However, the material properties that we extract will be applicable to fundamental questions in planetary science as discussed earlier, including small-body resurfacing and age estimates, which inform the dynamical population during the lifetime of the body. Legacy analysis of the plate penetrometer includes experience in modeling the compliance of Bennu's surface to the TAGSAM aboard OSIRIS-REx [22].

*5.2 Modeling granular dynamics*

Finding the material properties to numerically match the regolith response to vibrator activation at different centrifuging rates (Fig. 2) and at different locations within the chamber will give us excellent measurements of the restitution coefficients and cohesive properties of grains interacting at ultra-slow speeds. This occurs in the wake of impact events and during the reaccumulation process that forms the small rubble-pile bodies such as the Didymos components, Bennu, Ryugu, Phobos, and arguably comets such as Comet 67P/Churyumov–Gerasimenko [23] (although the current design of AOSAT+ will only be using material from one meteorite sample).

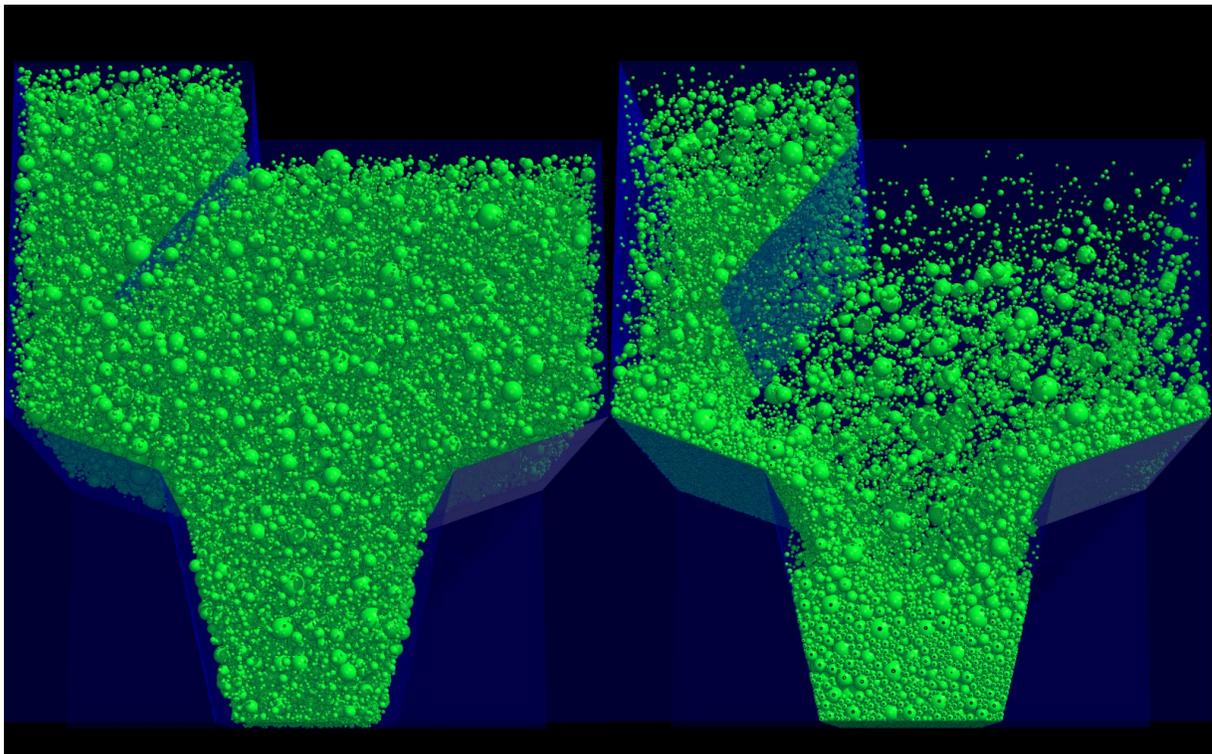

**Fig. 2**. Simulation of AOSAT+ payload spinning at 0.25 RPM. Simulation begins (*left frame*) with regolith occupying the entire volume (granular-gas phase), each grain with 1-cm/s dispersion. After minutes of collisional dissipation and the stratified centrifugal acceleration in the chamber (*right frame*), regolith begins to cascade (granular-liquid phase-transition) into the test-bed 'pool' below. Grain-size distribution follows power law ($d^{-3}$) and ranges from 1 mm to 1 cm (~100,000 particles). This simulation uses a coefficient of restitution (COR) of 0.8, though real values will likely be much closer to 1.0 [21]. Recording this behavior and matching to simulations will provide excellent constraints on regolith parameters including the COR, a fundamental mechanical property.

Simulations of grain-settling in the lab chamber under the conditions of 0.25-RPM rotation (Ryugu-like gravity) have been modeled. The full number and accurate size distribution of grains in the chamber have been modeled given certain assumptions on the entropy of the system (which will be constrained by the experimental data). Using different restitution parameters produces different settling times and size-stratification in the settled medium (Fig. 2). We are prepared to match these simulations with the experimental readings to extract these properties.

*5.3 Simulation of gas entrainment*

The effects of the nitrogen-gas injection will be observed and effects implicitly modeled in our granular code in order to extrapolate these effects to other regimes such as the TAGSAM head during sample-acquisition on Bennu. Explicit models of fluid-grain interactions is complex, but steps to include these effects in our code are underway in our funded collaboration with l'Ecole des Mines de Paris working with their Finite-Element code modelers [24]. However, simulating implicit effects (external grain-forcing and/or inclusion of small, high-speed particles) will increase our ability to extrapolate the effects of the gas jet beyond what we will achieve from the experimental data alone.

*5.4 Bead-deployer and ballistic penetration modeling*

Our team has successfully modeled a low-speed impact experiment performed to test the sampling mechanism aboard Hayabusa and Hayabusa2 [25,10]. That work placed constraints on the parameter space of the material properties of the glass beads involved in that experiment (performed in Earth-gravity) and confirmed the dependency on impactor shape. We also discovered how specific material properties affect the amount of material ejected. This will be applied to the case of AOSAT+ and the small centrifugal acceleration conditions of each experiment.

*5.5 Spacesuit material properties*

Observations of regolith motion and angle-of-repose on top of the four spacesuit-material coupons will likewise be matched in simulation to help determine their material properties. We have experience in measuring the angle-of-repose and its dependence on the physical and material properties of the system [21].

## 6. Conclusions

The AOSAT+ mission is an on-orbit centrifuge laboratory that will be used to create long-term, stable, low-gravity fields in which basic granular experiments can be performed and applied to the field of planetary science. Essential material properties of regolith behavior in low-gravity regimes will be extracted by using precise numerical simulations to reproduce these experiments. Supported by ground-based meteorite analysis, findings will help to characterize the physics, interaction behavior, and near-surface mechanics of small-bodies including Didymos-B, Bennu, Ryugu, Deimos and Phobos. The knowledge gained will help validate proposed mechanisms and mitigate risk for the next phase of small planetary body exploration, resource utilization, and hazardous asteroid mitigation. This approach to simulate low-gravity conditions provides a persistent link to off-world environments. By recreating these surface environments, we can test and validate robotic landing technology and eventually human adaptation to these environments and broaden our understanding and prove the feasibility of risky off-world surface exploration techniques before going to these locations.

**Acknowledgements**
This work was supported in part by NASA from Grant no. 80NSSC18K0226 as part of the OSIRIS-REx Participating Scientist Program. S.R.S. acknowledges support from the Université Côte d'Azur's Initiative d'EXcellence "Joint, Excellent, and Dynamic Initiative" (IDEX JEDI) of the Université Côte d'Azur.